\documentclass[twocolumn,pre,showpacs]{revtex4}
\usepackage{amssymb, amsmath, graphicx}
\def\be{\begin{equation}}
\def\ee{\end{equation}}
\begin{document}
\title{ Anomalous diffusions induced by enhancement of memory
  }
\author{Hyun-Joo \surname{Kim}}\email{hjkim21@knue.ac.kr}
\affiliation{
Department of Physics Education, Korea National University of Education,
Chungbuk 363-791, Korea\\
}

\received{\today}
\begin{abstract} 
We introduced simple microscopic non-Markovian walk models which describe underlying 
mechanism of anomalous diffusions. In the models, we considered the competitions
between randomness and memory effects of previous history
by introducing the probability parameters. The memory effects were considered in two aspects,
one is the perfect memory of whole history and the other is the latest memory enhanced with time.
In the perfect memory model superdiffusion was induced with the relation 
the Hurst exponent $H$ to the controlling parameter $p$ as $H=p$ for $p>1/2$. 
While in the latest memory enhancement models,
anomalous diffusions involving both superdiffusion and subdiffusion were induced with the relations
$H=(1+\alpha)/2$ and $H=(1-\alpha)/2$ for $0 \le \alpha \le 1$ where $\alpha$ is the parameter 
controlling the degree of the latest memory enhancement. Also we found that although the latest 
memory was only considered, the memory improved with time results in the long-range correlations
between steps and the correlations increase as time goes. Thus we suggest the memory enhancement
as a novel key origin describing anomalous diffusions.
\end{abstract}

\pacs{05.40.Fb, 02.50.Ey, 05.45.Tp}

\maketitle
\section{Introduction}
Random walks \cite{rw1} have played a key role in statistical physics for over a century.
They were proposed to stochastically formulate transport phenomena and macroscopic diffusion 
observables were calculated in long-time and short-distance limits of them \cite{rw2}.
It is well known that the key quantity characterizing the random walks or diffusion phenomena,
the mean squared displacement (MSD) $\langle x^2 (t) \rangle$, grows linearly with time. 
However, Hurst found the persistence of hydrologic time
series indicating that the MSD behaves in nonlinear way \cite{hurst,river,hjkim} and in recent,    
such phenomena have been observed in many different systems such as chaotic \cite{chaos}, biophysical
\cite{bio1, bio2, bio3,bio4,bio5}, economic systems\cite{econo1,econo2}, and etc.
The nonlinear behavior is recognized as anomalous diffusions compared with the linear behavior that
is regarded as normal diffusion, and is characterized in terms of the MSD 
\be
\langle x^2 (t) \rangle \sim t^{2H}.
\label{2h}
\ee
Here $\langle \cdots \rangle$ means average over independent realizations, i.e., ensemble
average, in general, in non-equilibrium.
$H$ is called as the anomalous diffusion or the Hurst exponent which classifies superdiffusion
($ H > 1/2 $) in which the past and future random variables are positively correlated and thus 
persistence is exhibited, and subdiffusion ($ 0 < H < 1/2 $) which behaves in the opposite way,
showing antipersistence.

The Hurst exponent however, does not provide any informations on underlying physical mechanism of  
anomalous diffusion, and so a variety of models to describe the mechanism 
have been proposed \cite{fbm,levy1,ctrw,Bou,havlin} but they do not give any a universal mechanism 
but rather suggest very distinct origins, separately. 
The representative models among them are the fractional Brownian motion (fBM) \cite{fbm}, 
the L\'{e}vy flights \cite{levy1,levy2,levy3,ks}, and the continuous time random walks (CTRW) \cite{ctrw,haus,ks}.
In the fBM, long-ranged temporal correlations between steps is given so that MSD scales like Eq. (\ref{2h})
within the range of $0<H<1$, and thus fBM describes both subdiffusion and superdiffusion however,
its correlation is mathematically constructed and it shows
stationary behaviors unlike nonstationary nature shown in real experiments and systems.
Meanwhile other two models mimic further specific systems and describe only one region of anomalous 
diffusions, respectively.
In L\'{e}vy flights, step-length distribution follows the power-law asymptotic behavior, so that the 
average distance per a step is infinite, which invokes superdiffusions. 
In CTRW model a time interval between two consecutive steps is a continuous random variable
which is drawn according to the waiting time distribution (WTD). For the WTD possessing the 
finite average of waiting time the MSD is linearly dependent on time, that is, the normal diffusive 
behavior is shown, while for the cases where the WTD behaves asymptotically as power-laws and thus
possesses infinite average of waiting time, subdiffusive behaviors are induced.
Also CTRW and L\'{e}vy walks have been generalized to reflect more physical realities 
by considering coupled space-time memory or various correlations between steps 
\cite{models,lebo, weiss, kutner1,kutner2, maso1, maso2}.
 
In recent years, a microscopic non-Markovian model with perfect memory of previous history
was proposed, in which a walker jumps persistently or antipersistently according to prior steps with
a probability parameter \cite{pre70}. Below the critical value of the control parameter, 
the model shows normal
diffusive behaviors while above it, superdiffusive behaviors. Due to its simpleness, the microscopic memory 
effect, the novel key origin of anomalous diffusion,  was easily applied to other models,
among which Cressoni et.al. suggested that the loss of recent memory rather than the distant past can induce
persistence, which is relate to the repetitive behaviors, psychological symptoms of Alzheimerś disease 
\cite{alzh}. 
In \cite{miad}, it was shown that by adding a possibility that a walker does not move at all in the model 
of \cite{pre70}, diffusive, superdiffusive, and subdiffusive behaviors can exhibit in different parameter 
regimes. It has advantage to describe the anomalous diffusion within a single model just by changing
the parameters, however, in this case, the subdiffusive property may be caused by the staying behavior rather
than the memory effect and thus superdiffusion and subdiffusion are not induced by a single origin.

Thus although anomalous diffusions have been described by various origins separately, more general origins
which can describe the nonstationary mechanisms in both superdiffusions and subdiffusions 
are still questionable. To answer this, we focus on two features,  
microscopic memory effect varying with time and the competition between Markovian and non-Markovian processes
which are realized by simple stochastic models.
In the first model, non-Markovian processes induced by the full memory of entire history and Markovian processes
constructed by the original random walk are competed by a probability parameter. 
In the second model, non-Markovian processes are induced by the latest memory rather than full memory and
its realizations vary with time. From these models we find that in the regime where nonMarkovian nature 
prevails, superdiffusion is induced by the perfect memory, while the latest memory enhanced with time cause 
subdiffusions as well as superdiffusions.

\section{Model with perfect memory}
First, we define a simple microscopic non-Markovian model in which a walker moves depending on full memory 
of its entire history with probability $p$ and at random with probability $1-p$. 
The random walker starts at origin and randomly moves either one step to the right or left at time $t = 1$, 
so the position of the walker becomes $x_1 = \sigma_1 $ with $\sigma_1 = 1 $ or -1. 
Then the random variable $\sigma_1$ is preserved in the set $\{\sigma \}$ to memory the entire history of 
the walking process. 
At time $t$, the stochastic evolution equation becomes as 
\be
x_{t+1} = x_{t} + \sigma_{t+1},
\label{x}
\ee
with
\be
\sigma_{t+1} = \left \{
\begin{array}{ll}
\sigma_{t'},  & \quad \text{with probability $p$}\\
+1 \; \text{or} \; -1,  & \quad \text{with probability $1-p$.}
\end{array} \right.
\ee
Here $t' \leq t$ and the random variable $\sigma_{t+1}$ is chosen from the set $\{\sigma_t'\}$ 
with equal probability $1/t$. For the case of probability $1-p$, $\sigma_{t+1}$ is chosen in 1 or -1 
with equal probability 1/2 at random. It differentiates this model from that 
of \cite{pre70} where $\sigma_{t+1} = -\sigma_{t'}$ which makes competitions between positive 
correlation of random variables and negative correlation rather than randomness in the process. 

 \begin{figure}[ht]
 \includegraphics[width=9cm]{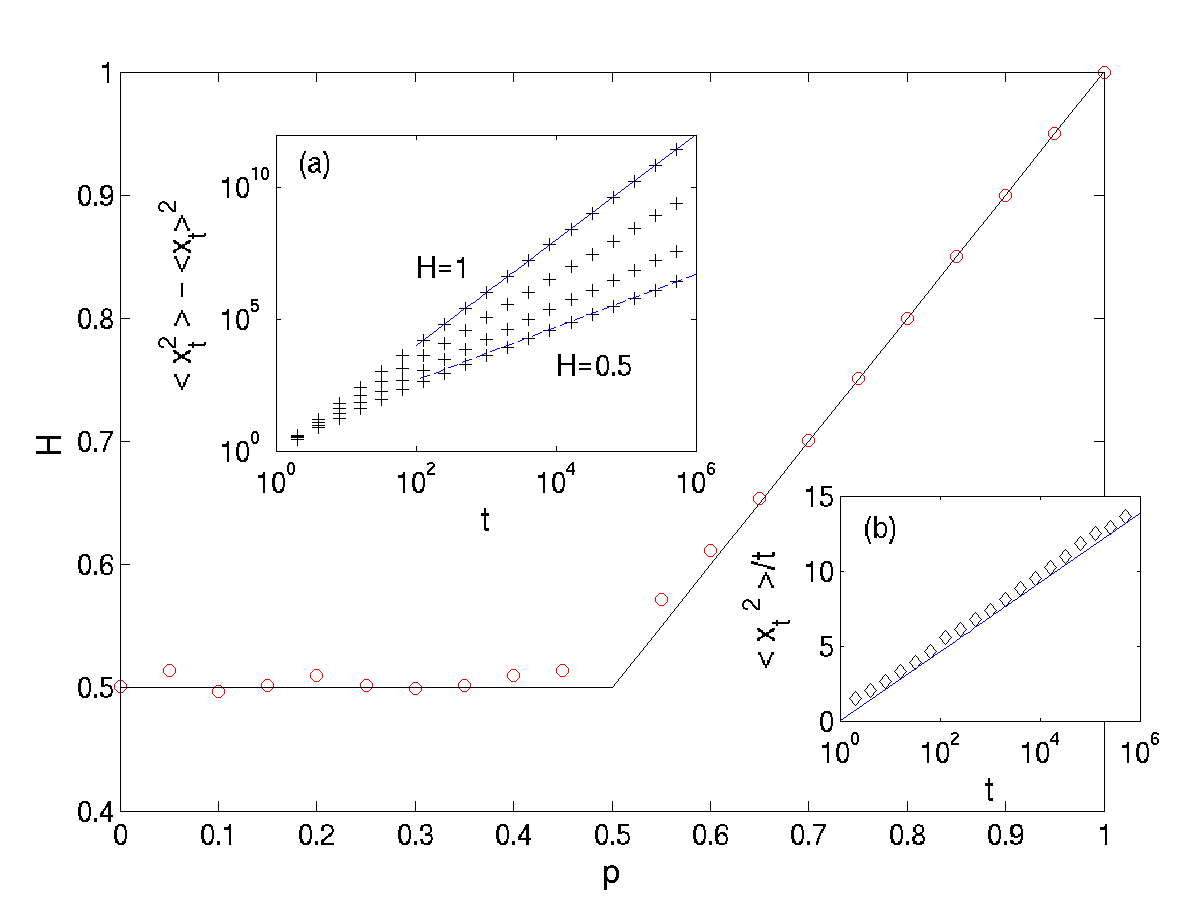}
 \caption{
 The inset (a) shows the plots of the variance $\langle x^2_t \rangle - \langle x_t \rangle ^2$ 
 for $p=0.4, 0.6, 0.8$, and 1 from the bottom to
 the top. For $p=1$, the data are in a excellent agreement with the solid line 
 indicating $H=1$, while for $p=0.4$, the data are in a good agreement with the dashed line 
 $H=0.5$. 
 The data were measured with the initial condition where a walker moves to the right or left with equal 
 probability $1/2$ and $10^4$ independent realizations.
 The main plot shows the Hurst exponent $H$ versus the parameter $p$. 
 The solid line is $H=0.5$ for $p<0.5$ and 
 $H=p$ for $p>0.5$. It confirms the analytic results of Eq. (\ref{x2a1}) and (\ref{x2a3})
 which shows that the persistence vanished in the regimes $p<0.5$  and there are the persistence with 
 the relation, $H=p$ for $p>0.5$.
 The case of $p=0.5$ is shown in the inset (b) which shows the marginal behavior, 
 $\langle x^2_t \rangle /t$ increases logarithmically.
 }
 \label{hltm}
 \end{figure}
In order to compute the mean displacement $\left< x_t \right >$, we first note that for a given the previous
history $\{\sigma_t \}$, the conditional probability that $\sigma_{t+1} = \sigma$ can be written as
\be
P[\sigma_{t+1} = \sigma | \{\sigma_t \}] \;= \;\frac{1-p}{2} \;+\; p\frac{t_\sigma}{t}\; =\; \frac{1}{2t} \;\sum_{k=1}^ t 
\;(\: p\:\sigma_k \sigma \;+\; 1),
\label{P}
\ee
where $t_\sigma$ is the total number of steps having $\sigma$ in the past.
For $t \geq 1 $ the conditional mean value of $\sigma_{t+1}$ in a given realization is given by
\be
\left<\sigma_{t+1} | \{ \sigma_t \} \right> = \sum_{\sigma = \pm 1} \sigma P [\sigma_{t+1} = \sigma | \{\sigma_t \} ]
= \frac{p}{t} x_t,
\label{sigma}
\ee 
where the displacement from the origin becomes as $x_t = \sum_{k=1}^t \sigma_k$ if the walker starts at $x=0$. 
On averaging Eq. (\ref{sigma}) over all realizations of the process, the conventional mean value of
$\sigma$ is given by
\be
\left<\sigma_{t+1}\right> = \frac{p}{t} \left< x_t \right>.
\label{asigma}
\ee
and by using the average of Eq. (\ref{x}) the recursion relation is obtained as
\be
\left< x_{t+1} \right>=\left( 1 + \frac{p}{t} \right) \left< x_t \right >.
\label{recursion}
\ee
The solution of Eq. (\ref{recursion}) is given as 
\be
\left<x_t\right> =\left<\sigma_1\right>\frac{\Gamma (t+p)}{\Gamma (1+p) \Gamma (t) } \sim t^p, \text{  for $t \gg 1$}.
\label{xt}  
\ee
When $\langle \sigma_1 \rangle \neq 0$ the mean displacement increases monotonically following the power-law.
It is the same behavior as that of \cite{pre70} although the exponent is different, which indicates that the persistent due to the full memory makes walker moves away from the origin with time on average. 

\begin{figure}
\includegraphics[width=9cm]{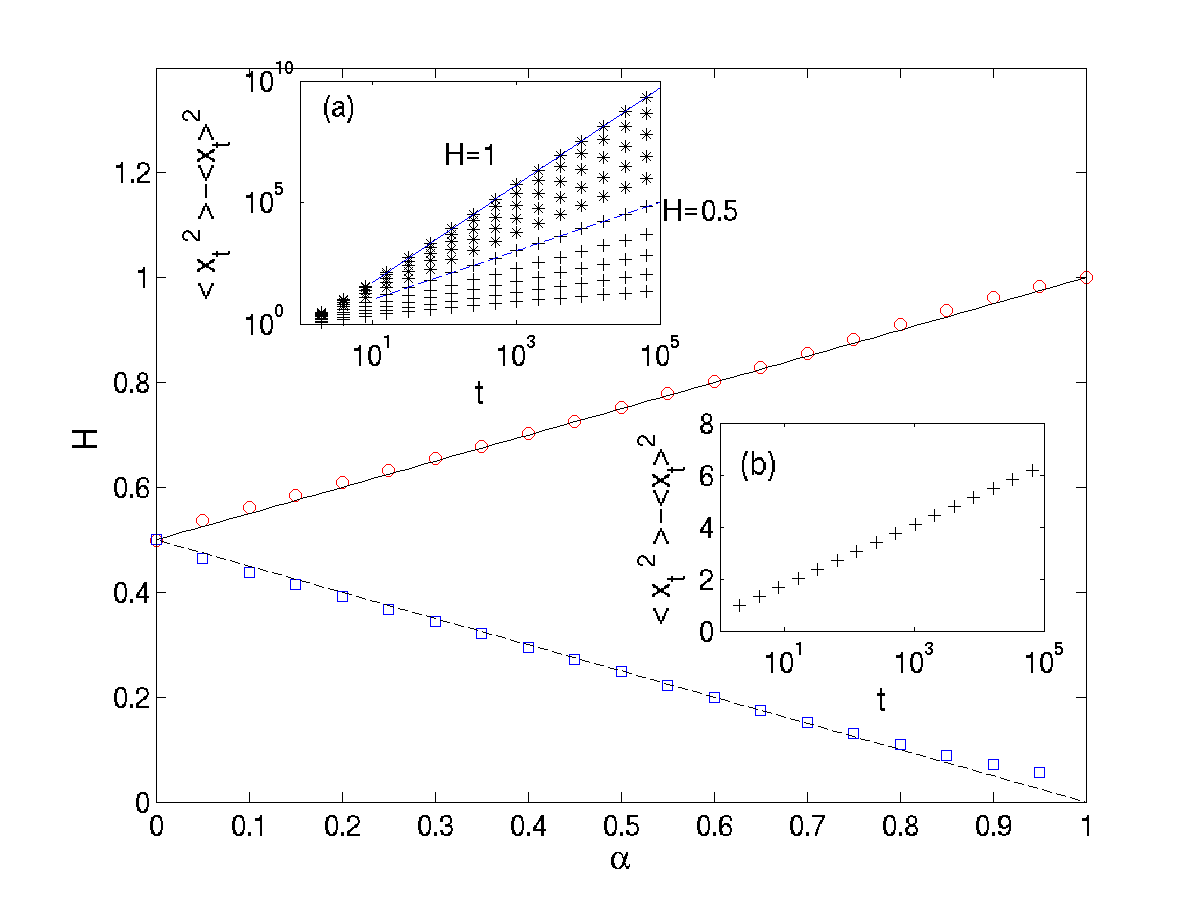}
\caption{The inset (a) shows the plots of $\langle x^2_t \rangle- \langle x_t \rangle ^2$ 
with $\alpha=0.2, 0.4, 0.6, 0.8$, and 1 
for the pLMEM (star symbols) and with $\alpha=0, 0.2, 0.4, 0.6,$ and 0.8 for
the nLMEM (plus symbols).
For the models, the mean displacement is nearly zero numerically so that the variance is equal to the
MSD.
The solid and dashed lines represent the case of $H=1$ and $H=0.5$, respectively.
The plot shows the Hurst exponent $H$ as a function of the parameter $p$ measured with $5 \times 10^5$ 
independent realizations. The circle symbols and the square symbols represent the data for the pLMEM
and the nLMEM, respectively. The solid line is $H=(1+\alpha)/2$ and the dashed line is $H=(1-\alpha)/2$.
It is shown that the data are in a good agreement with the lines, which indicates the LMEMs
can well describe the all anomalous diffusions including superdiffusion and subdiffusions.
The case of $\alpha=1$ for the nLEME is shown in the inset (b) which shows the marginal behavior, 
$\langle x^2_t \rangle- \langle x_t \rangle ^2$ increases logarithmically.}
\label{hstm}
\end{figure}
The recursion relation of the MSD also can be computed from Eq. (\ref{P})
and Eq. (\ref{x}) as follows:
\be
\left< x_{t+1}^2 \right>=1+ \left( 1 + \frac{2p}{t} \right) \left< x_t^2 \right >,
\label{x2}
\ee
and its solution \cite{pre70} is asymptotically obtained as 
\be
\left< x_{t}^2 \right>=\frac{t}{1-2p},  \quad\text{ for $p<1/2$},
\label{x2a1}
\ee
\be
\left< x_{t}^2 \right>= t \:\mathrm {ln} \:t, \quad \text{ for $p=1/2$},
\label{x2a2}
\ee
\be
\left< x_{t}^2 \right>=\frac{t^{2p}}{(2p-1)\Gamma (2p)},  \quad\text{ for $p>1/2$}.
\label{x2a3}
\ee
For $p<1/2$, the MSD depends linearly on time and the mean displacement follows the power-law 
$\left<x_t \right> \sim t^{p}$ with the exponent smaller than $1/2$, so that 
the variance $\Delta (t) = \left< x_t ^2 \right> 
- \left< x_t\right>^2 $ remains normally diffusive for large $t$. Specially when at $t=1$ a walker moves to
the right or left with equal probability $1/2$ so that $\left< \sigma_1 \right>=0$, 
the variance increases asymptotically linearly with time having the diffusion coefficient $D=1/2(1-2p)$.
While for $p>1/2$ the MSD follows the power-law $\left<x_t^2 \right> \sim t^{2p}$ which is of the
same order as the square of the mean, but with a different prefactor. 
Hence it results in the superdiffusion with the relation between the Hurst 
exponent and the parameter $H=p$. 
The marginal superdiffusive phenomena is shown for $p=1/2$. 
These results have been confirmed by computer simulations as shown in the Fig. \ref{hltm}.
The critical parameter $p_c = 1/2$ means that the superdiffusive phenomena occur when
the persistence induced by full memory prevails in the process against the randomness.
It can be compared to the results of \cite{pre70} in which the critical value of the parameter ($p_c = 3/4$)
is larger than that of this model, which indicates that the antipersistent rule invokes more randomness in the 
process than just random choices used in this model. The steps made by anticorrelation with previous steps
do not continuously retain antipersistent nature but rather bring about random nature changing the directions 
of steps. Therefore it is difficult to embody subdiffusion phenomena from the perfect memory effect and thus
we need to consider a novel approach to describe anomalous diffusions comprising subdiffusions.

\section{ Models with memory enhancement}

We suggest the following new non-Markovian stochastic model
where for $t > 1$, $\sigma_{t+1}$ is given by
\be
\sigma_{t+1} = \left \{
\begin{array}{ll}
\sigma_{t},  & \text{with probability $1-1/t^{\alpha}$}\\
1 \; \text{or} \; -1,  & \text{with probability $1/t^{\alpha}$}
\end{array} \right.
\label{stm}
\ee
and the walker starts at origin and moves to the right or left with equal probability at time $t=1$.
Over time, the probability of taking the same direction with the latest step increases
and the larger value of parameter $\alpha$ is, the much faster the probability grows with time. 
That is, in this model only the latest step is remembered unlike the above perfect memory model,
and the persistence with the previous step is enhanced with time of which degree is 
controlled by the parameter $\alpha$.
When $\alpha =0$ it reduced to the original random walk.
We shall refer to this model as the positive latest memory enhancement model(pLMEM). 
Meanwhile in Eq. (\ref{stm}) if the rule $\sigma_{t+1} = -\sigma_{t}$ is taken the correlation between two
successive steps is negative and thus let's call this the negative latest memory enhancement model (nLMEM).

The computer simulations were run for these two LMEMs. Figure \ref{hstm} shows 
the Hurst exponent $H$ versus the parameter $\alpha$ for the pLMEM (circles) and for
the nLMEM (squares). The solid line represents that the Hurst exponent $H$ relates to the parameter $\alpha$
as $H = (1+\alpha)/2$ for $0 \leq \alpha \leq 1$ for the PLMM. 
For the case of $\alpha > 1$, the probability $p(t) = 1-1/t^{\alpha} $ approaches to one 
so fast than the case of $\alpha = 1$ as time becomes large, so that it also shows ballistic motions resulting
in $H=1$.
While the dashed line represents that 
$H = (1-\alpha)/2$ for the NLMM. For the case $\alpha = 1$ of the nLMEM it shows the marginal behavior
showing $ \left<x^2_t \right> - \langle x_t \rangle ^2 \sim \mathrm{ln} t $ (the inset (b) in Fig. \ref{hstm}).
Thus the LMEMs well brought about both superdiffusions and subdiffusions with a single origin, 
although considering only for the latest memory. 
It is compared that a walk process just depending on short term memory at each time 
is reduced into the original random process. Therefore it can be regarded as a new nonstationary microscopic mechanism describing anomalous diffusions.

 \begin{figure}[ht]
 \includegraphics[width=9cm]{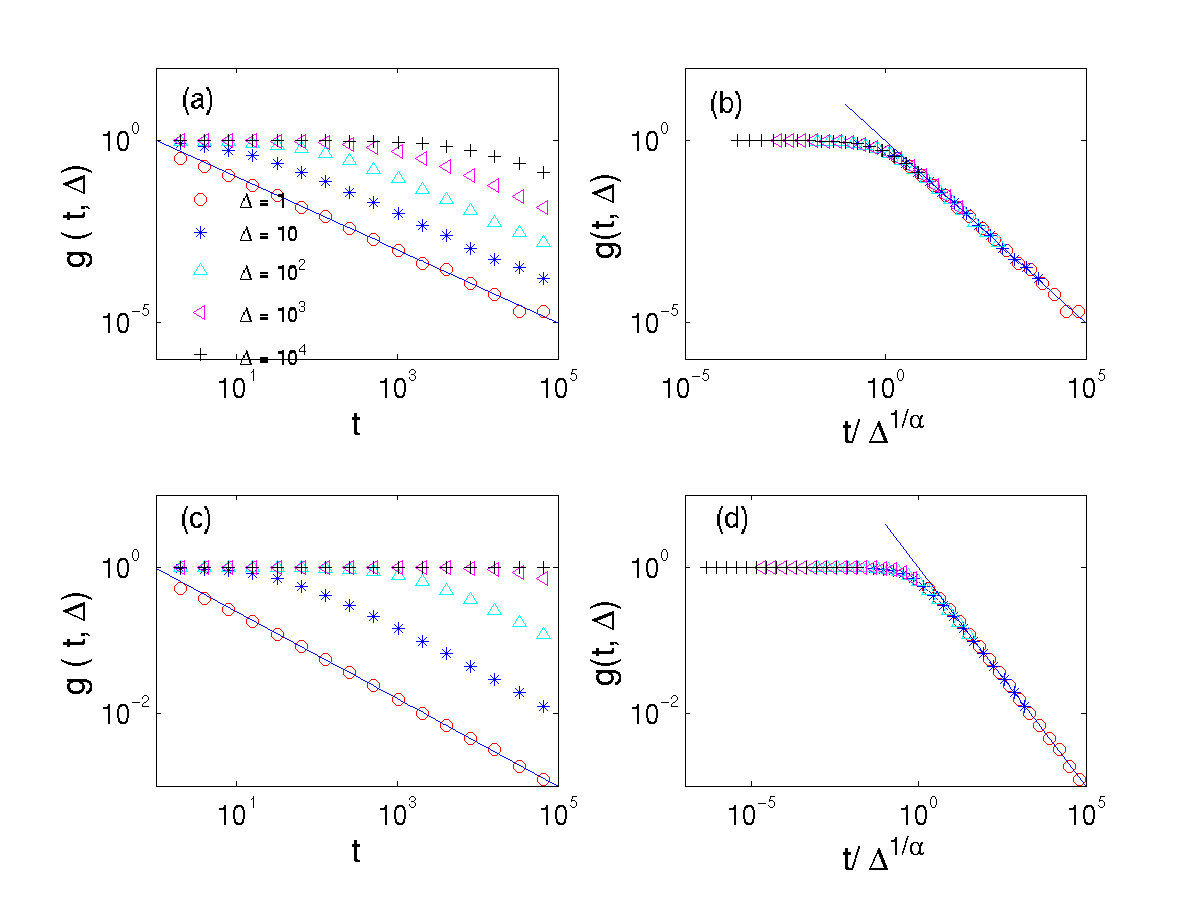}
 \caption{
 (a) The function $g(t, \Delta)$ is shown as a function of time $t$ for various
 intervals $\Delta = 1, 10, 10^2, 10^3,$ and $10^4$ for $\alpha = 1$. The solid line represents 
 $g(t,1) \sim t^{-\alpha}$. (b) The functions $g(t, \Delta)$ are fallen into a single curve by scaling
 time as $t/\Delta^{1/\alpha}$. The solid line represents that $g(t,\Delta) \sim \Delta/ t^{\alpha}$.
 (c) The function $g(t, \Delta)$ for $\alpha = 0.6$. (d) The data are well collapsed with 
 the critical time $t_c = t/\Delta^{1/\alpha}$ for $\alpha=0.6$,.
 }
 \label{corr_super}
 \end{figure}

\section{time-varying Correlations} 
In order to study in detail how the memory enhancement affects to the correlations between steps
we consider the correlation function $C(t, \Delta)$ defined as 
\be
C(t, \Delta) = \left<\: \sigma_t \;\sigma_{t+\Delta}   \:\right>-\left<\: \sigma_t \:\right>\left<\: \sigma_{t+\Delta}   \:\right>
\ee 
where when $\Delta = 1$, $C(t, 1) \sim 1 - t^{-\alpha}$ may be given by the Eq. (\ref{stm})
and $\langle \cdots \rangle$ is the average for independent realizations.
For convenience, we considered a function $g(t, \Delta) \equiv 1 - C(t, \Delta)$ and measured it for 
different values of $\Delta$. Figure \ref{corr_super} (a) shows the function $g(t, \Delta)$ versus time $t$ 
for various values of $\Delta$ for $\alpha = 1$. 
The solid line represents that $g(t, 1) \sim t^{-\alpha}$ for $t>1$ 
with $\alpha=1$ as expected.
Meanwhile it shows that for $\Delta > 1$, the function $g$ becomes 1 
for $t \ll t_c$ and $g(t) \sim t^{-\alpha}$
for $t \gg t_c$. The data collapse into a single curve very well with $t_c = \Delta^{1/\alpha}$ as shown in the 
Fig. \ref{corr_super} (b). Figure \ref{corr_super} (c) and (d) show the same results for $\alpha = 0.6$. 
Thus the correlation function $C(t, \Delta)$ scales as 
\be
C(t, \Delta) \sim  \left \{
\begin{array}{ll}
\text{0},  & \quad \text{for $t \ll \Delta^{1/\alpha}$}\\
1 - \Delta / t^\alpha,  & \quad \text{for $t \gg \Delta^{1/\alpha.}$}
\end{array} \right.
\label{cfunction}
\ee
At the critical time after which the correlations appears, the persistent probability to follow the last
step is $p(t_c) = 1-1/\Delta$. Although the present step just only depends on only one preceding step
, it generates the correlations between steps far away from each other when the persistent probability is
larger than the critical probability $p(t_c)$. That is, the shortest term memory increasing with time can 
induce the long-range correlations in enough long time limits. Also it has to be addressed that 
unlike the stationary series of the fBM in which the correlation does not change with time and only depend on a
time interval like as $C(\Delta) \sim \Delta^{-2(1-H)}$, this process is nonstationary and the correlations 
depend on time interval as well as time $t$. In the fBM the correlations decrease as the  
interval increases depending on the Hurst exponent $H$, while the correlation in the pLMEM decreases linearly
with the interval irregardless of $H$ and increases over time. The larger value of $\alpha$ is
the much faster the correlation increase and so the more superdiffusive behaviors appear. 

 \begin{figure}[ht]
 \includegraphics[width=9cm]{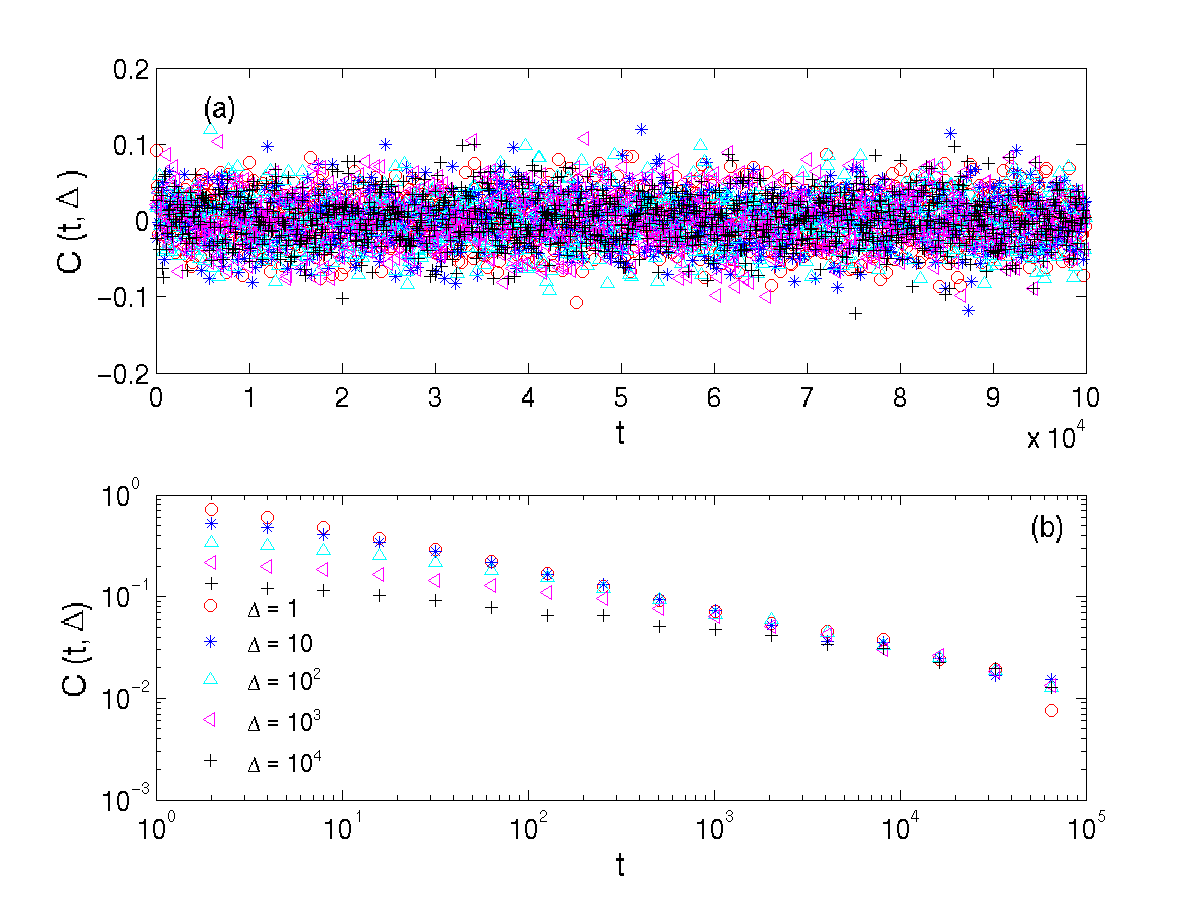}
 \caption{
 (a) The correlation function $C(t, \Delta)$ as a function of time $t$ for various intervals
 $\Delta = 1, 10, 10^2, 10^3, and 10^4$ for $p=0.2$ of the perfect memory model.
 In the case, the normal diffusion is shown and thus there are no correlations between steps.
 (b) The correlation function $C(t, \Delta)$ for $p=0.8$ of the perfect memory model.
 The correlations between steps are positive for all measured times and intervals as expected
 in superdiffusions. However they show the dependence on the time as well as the interval unlike
 the stationary process like the fBM.
 }
 \label{corr_ltm}
 \end{figure}
 \begin{figure}[ht]
 \includegraphics[width=9cm]{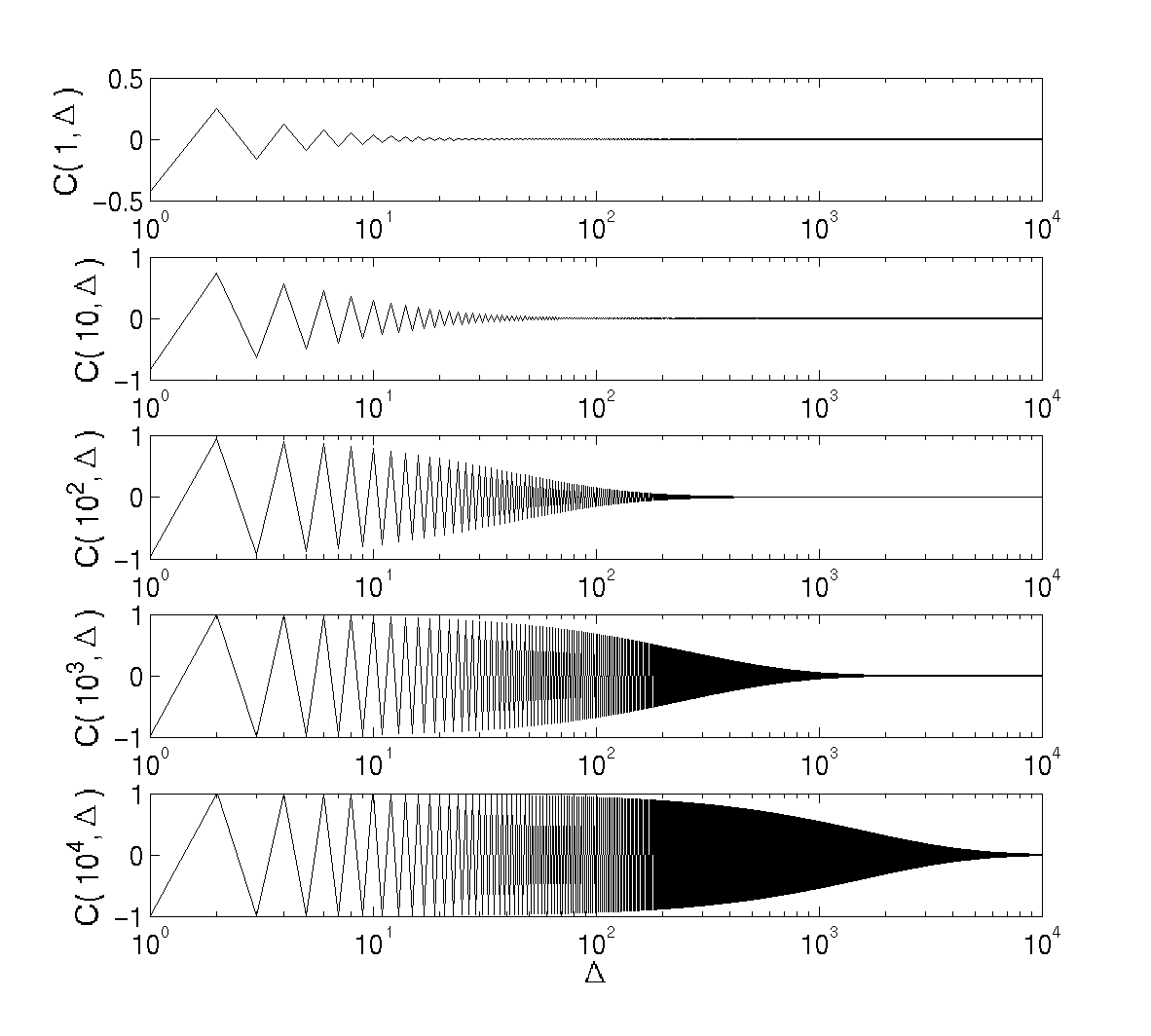}
 \caption{
 The semi-log plot of the correlation functions $C(t_\times, \Delta)$ as a function of interval 
 $\Delta$ at different times $t=1, 10, 10^2, 10^3,$ and $10^4$ for $\alpha = 0.8$ of the nLMEM.
 }
 \label{corrsub}
 \end{figure}

These properties of the correlation function are distinguished from those of the perfect memory model
in which the correlation function decreases when time goes. As shown in the Fig. \ref{corr_ltm} (a), 
for $p=0.2$ the correlation function $C(t, \Delta)$ does not depend on time $t$ as well as interval 
$\Delta$ and their averages for time  become zero, which represents the normal diffusion. 
For $p=0.8$ (Fig. \ref{corr_ltm} (b)) the steps are positively correlated as expected to make the 
super diffusions. The correlation functions decrease when the time interval $\Delta$ becomes large 
at same time and the difference lessens with time.
That is, the process is also nonstationary process and the correlation function depends on
the time however, it decreases with time unlike the pLMEM. Thus the perfect memory of whole history and the
latest memory increasing with time are two different origins resulting in superdiffusive behaviors.

For the nLMEM, the time dependency of the absolute correlation function is the same as the pLEME,
because of the probability following the latest step is same for two LMEMs
irrespective of the given sign, positive or negative correlation. For the nLMEM the correlation function 
as a function of interval $\Delta$ may be negative or oscillatory due to the subdiffusive nature. 
Considering nonstationary behaviors of the process, we measured the correlation function
at fixed time $t_\times$ 
given by $C(t_\times, \Delta) = \langle \sigma_{t_\times} \sigma_{t_\times + \Delta} \rangle 
- \langle \sigma_{t_\times}\rangle \langle \sigma_{t_\times + \Delta}\rangle$.
Figure \ref{corrsub} shows the correlation function as a function of interval $\Delta$ at different 
times. $C(t_\times, \Delta)$ oscillates totally in the nonzero regimes, which is distinguished from the
subdiffusions of the fBM with negative correlation like as $C(\Delta) \sim -\Delta^{-2(1-H)}$. 
Also, when $t_\times$ becomes large the oscillatory range is more longer. 
That is, like Eq. (\ref{cfunction}), $C(t_\times, \Delta)$ becomes as 
\be
|C(t_\times, \Delta)| \sim  \left \{
\begin{array}{ll}
1 - \Delta / t_\times^\alpha,  & \quad \text{for $\Delta \ll t_\times^{\alpha}$}\\
\text{0},  & \quad \text{for $\Delta \gg t_\times^{\alpha}$}.
\end{array} \right.
\label{gfunction}
\ee 
It indicates that the longer time is, the more anticorrelation is persistent.

\section{conclusion}
In conclusion, the microscopic nonstationary mechanisms of anomalous diffusions 
have been studied through the simple new models with memory effects of previous walk processes. 
In the models, Markovian and nonMarkovian processes were
controlled by the probability parameter, and anomalous diffusions were induced with the Hurst 
exponent related to the parameter. The perfect memory of whole history invokes superdiffusive
behaviors with $H=p$ for $p>0.5$ in which regime the nonMarkovian nature prevails, 
while subdiffusions are not invoked.
The anomalous diffusive behaviors involving both superdiffusions and subdiffusions
could be described in the mechanism where the latest memory increases with time. 
The persistent behaviors with the latest memory enhancement induced the superdiffusions
with $H=(1+\alpha)/2$. While taking the opposite direction to the latest step brought about the 
subdiffusive behaviors with $H=(1-\alpha)/2$. 
The perfect memory resulted in the long-range step correlation decreasing with time, while 
even though the memory is restricted to the latest step, the memory enhancement resulted in the 
long-range correlations increasing with time above the critical time which increases with the interval.
Thus the enhancement of memory may be a novel key origin describing all anomalous diffusions
and we expect that these time-varying features will be measured in various real systems showing 
anomalous diffusions and these simple models can be served as basic models in studying another various 
aspects of anomalous diffusive phenomena.

Meanwhile we need to consider the ergodicity breaking in the processes. It is known that nonstationary 
stochastic processes are generally not ergodic, that is, the means as ensemble averages are different from
those as time averages \cite{ergo1,ergo2,ergo3,ergo4,ergo5}. 
These processes are nonstationary due to the memory effect and all 
the analysis in this study was made on the basis of nonequilibrium ensemble averages, so that  
it does not mean that when time average is run it gives the same Hurst exponent as the above provided  
. We are going to deal with the ergodicity breaking in these models in elsewhere.

\end{document}